\documentstyle[12pt,a4wide]{article}
\newcommand{\be}{\begin{equation}}
\newcommand{\ee}{\end{equation}}
\newcommand{\bea}{\begin{eqnarray}}
\newcommand{\eea}{\end{eqnarray}}
\newcommand{\bean}{\begin{eqnarray*}}

\newcommand{\eean}{\end{eqnarray*}}
\font\upright=cmu10 scaled\magstep1
\font\sans=cmss10
\newcommand{\ssf}{\sans}
\newcommand{\stroke}{\vrule height8pt width0.4pt depth-0.1pt}
\newcommand{\Z}{\hbox{\upright\rlap{\ssf Z}\kern 2.7pt {\ssf Z}}}

\newcommand{\C}{{\rlap{\rlap{C}\kern 3.8pt\stroke}\phantom{C}}}
\newcommand{\R}{\hbox{\upright\rlap{I}\kern 1.7pt R}}
\newcommand{\CP}{\C{\upright\rlap{I}\kern 1.5pt P}}
\newcommand{\PP}{\hbox{\upright\rlap{I}\kern 1.5pt P}}

\newcommand{\identity}{{\upright\rlap{1}\kern 2.0pt 1}}

\newcommand{\hf}{\frac{1}{2}}
\newcommand{\pp}{\Delta}
\newcommand{\HH}{\mbox{\hbox{\upright\rlap{I}\kern 1.7pt H}}}
\newcommand{\kk}{\kappa}

\newcommand{\fr}{\frac}
\newcommand{\lm}{\lambda}

\newcommand{\pr}{\partial}
\newcommand{\hs}{\hspace{5mm}}

\newcommand{\dg}{\dagger}

\newcommand{\ve}{\varepsilon}
\newcommand{\acc}{\\[3mm]}

\newcommand{\zb}{{\bar z}}

\newcommand{\news}{ }
\begin{document}
\title{\vskip -70pt
\begin{flushright}
{\normalsize UKC/IMS/99/07} \\
\end{flushright}\vskip 50pt
{\bf \large \bf Monopoles and Harmonic Maps}\\[30pt]
\author{Theodora Ioannidou and Paul M. Sutcliffe\\[10pt]
\\{\normalsize  {\sl Institute of Mathematics, University of Kent at Canterbury,}}\\
{\normalsize {\sl Canterbury, CT2 7NZ, U.K.}}\\
{\normalsize{\sl Email : T.Ioannidou@ukc.ac.uk}}\\
{\normalsize{\sl Email : P.M.Sutcliffe@ukc.ac.uk}}\\}}
\date{September 1999}
\maketitle

\begin{abstract}
\noindent Recently Jarvis has proved a correspondence between $SU(N)$  monopoles and
rational maps of the Riemann sphere into flag manifolds. Furthermore, he has
outlined a construction to obtain the monopole fields from the rational map.
In this paper we examine this construction in some detail and provide explicit
examples for spherically symmetric $SU(N)$ monopoles with various symmetry breakings.
In particular we show how to obtain these monopoles from harmonic maps into 
complex projective spaces. The approach extends in a natural way to  monopoles
in hyperbolic space and we use it to construct new spherically
 symmetric $SU(N)$ hyperbolic monopoles.
\end{abstract}

\newpage

{\Large {\bf I. Introduction}}\\

\news\ \ \ \ \ \
This paper is concerned with static $SU(N)$ BPS monopoles, which are topological
solitons in a Yang-Mills-Higgs gauge theory. The Bogomolny equation, which describes
all static monopoles, is integrable and so a variety of techniques are available
for studying monopoles, including twistor methods. Despite this fact there are still
only a limited number of known explicit monopole solutions, though the integrability
of the Bogomolny equation allows many features of monopoles, such as the dimensions
of their moduli spaces, to be determined. 

An example where the integrability of the Bogomolny equation can be used to prove
results on monopoles is the correspondence proved by Jarvis \cite{Ja}, between
monopoles and rational maps from the Riemann sphere into flag manifolds.
The rational map arises as the scattering data, along half-lines from the origin,
of a linear operator constructed from the monopole fields. Furthermore, in proving
the correspondence Jarvis outlines an \lq inverse scattering\rq\ procedure whereby
the monopole fields can be reconstructed from the rational map.
It is this construction which is the focus of this paper. The construction involves
solving a nonlinear partial differential equation which is equivalent to the
Bogomolny equation, but for which the boundary conditions are given in terms of the
rational map. This is the main point of the construction, since for the original Bogomolny
equation it is not at all clear how to specify boundary conditions on the fields so as
to obtain a unique monopole solution. We perform the construction explicitly
for several examples of $SU(N)$ monopoles with spherical symmetry and a variety of
symmetry breakings. The solutions are obtained from harmonic maps of the plane into
\CP$^{N-1}$, with the degrees of the harmonic maps related to the topological charges
 of the monopoles.

Perhaps we should make it clear at this point that there are several approaches
to studying spherically symmetric monopoles \cite{BW,BWe,GGO,BCGPS} and the main aim
of this paper is not the construction of new monopole solutions, but rather to gain
a better understanding of the correspondence between monopoles and rational maps.
In particular we study the construction of monopole solutions from the rational map data,
and $SU(N)$ monopoles with spherical symmetry are a good vehicle for this.

The construction of monopoles from rational maps has a natural generalization to 
monopoles in hyperbolic space. Using this approach we construct explicit  solutions 
for spherically symmetric $SU(N)$ hyperbolic monopoles. As far as we are aware
these multi-monopole solutions are new. As we shall see, the construction of hyperbolic
monopoles has a simplifying feature in comparison to the Euclidean case, and therefore
a useful way to obtain the Euclidean solutions is as the zero curvature limit of
the hyperbolic ones.\\
 
\newpage
{\Large{\bf II. SU(N) Monopoles}}\\

\news\ \ \ \ \ \
BPS monopoles are finite energy solutions to the Bogomolny equation
\be
D_i \Phi=-\fr{1}{2} \ve_{ijk}\,F^{jk}
\label{Bog}
\ee
where $D_i=\pr_i+[A_i, ]$ is the covariant derivative with $A_i$, for $i=1,2,3$, an
$su(N)$-valued gauge potential with gauge field $F_{jk}=\pr_j A_k-\pr_k
A_j+[A_j,A_k].$ 
The Higgs field, $\Phi$, is an $su(N)$-valued scalar field for which
nontrivial asymptotic boundary conditions are imposed so that
topological solitons exist.
More precisely, there is a choice of gauge such that in a given direction the 
Higgs field for large radius $r$ is given by
\be
\Phi=i\Phi_0-\fr{i}{r}\Phi_1+{\rm O}(\fr{1}{r^2})
\label{cond}
\ee
where $\Phi_0=\mbox{diag}(\lm_1, \lm_2, \dots,\lm_N)$, 
with the trace-free condition requiring that $\lm_1+\lm_2+\dots+\lm_N=0$, 
and we choose the ordering such that $\lm_1\geq\lm_2\geq \dots
\geq\lm_N$.
$\Phi_1$ is another diagonal matrix, 
$\Phi_1=\fr{1}{2}\mbox{diag}(n_1,
\,n_2-n_1,\,\dots,\,n_{N-1}-n_{N-2},\,-n_{N-1})$, and it can be shown that
the numbers $n_1$, $n_2$, ..., $n_{N-1}$ are always integers.
 
$\Phi_0$ is the vacuum expectation value of $\Phi$ and it breaks the symmetry group
from $SU(N)$ to a residual symmetry group $J$, given by the isotropy group of $\Phi_0.$
The Higgs field on the two-sphere at infinity defines a map from $S^2$ to the coset
space of vacua $SU(N)/J$, so that when $\pi_2(SU(N)/J)$ is non-trivial then all
solutions have a topological characterization.

If the $\lm_i$ are all distinct then the residual symmetry group is the maximal
torus, that is, $J=U(1)^{N-1}$, and this is known as maximal symmetry breaking.
In this case
\be
\pi_2\left(\frac{SU(N)}{U(1)^{N-1}}\right)=\pi_1(U(1)^{N-1})=\Z^{N-1}
\ee
so the monopoles are associated with $N-1$ integers, which are called the
topological charges, and are precisely the integers 
 $n_1$, $n_2$, ..., $n_{N-1}$ appearing in $\Phi_1.$

In contrast the case of minimal symmetry breaking is when all but one of
 the $\lm_i$ coincide, so the residual symmetry group is $U(N-1)$.
Since
\be
\pi_2\left(\frac{SU(N)}{U(N-1)}\right)=\Z
\ee 
there is only one topological charge in this case and the remaining
integers are called magnetic weights. The simplest way to distinguish
the topological charges from the magnetic weights is to examine the
expression for the energy of the monopole.

The condition (\ref{cond}) guarantees that the
configuration has finite energy
\be
E=\fr{1}{4\pi}\int
-\mbox{tr}\left(\fr{1}{2}\,F_{ij}^2+(D_i\Phi)^2\right)\,d^3x.
\ee
The energy depends only on the topological charges and the asymptotic
eigenvalues of $\Phi$, in fact 
\be 
E=(\lm_1-\lm_2)n_1+(\lm_2-\lm_3)n_2+\dots+(\lm_{N-1}-\lm_N)n_{N-1}.
\ee
From this expression it can be seen that the difference $\lm_j-\lm_{j+1}$
determines the mass of the monopole of type $j$, of which there are $n_j$ in
the given solution. In the minimal symmetry breaking case, where 
 $\lm_2=\lm_3=\dots=\lm_{N}$, then all but the first type of monopole becomes
massless, so that $n_1$ remains a topological charge but the remaining integers
become magnetic weights and do not contribute to the value of the energy.
Note that we can not distinguish in a gauge invariant way between $n_2-n_1$, ...,  
$n_{N-1}-n_{N-2}$ and $-n_{N-1}$, and so when we refer to values of magnetic weights
it is understood that this equivalence should be applied.

For intermediate cases of symmetry breaking the residual symmetry group is
$J=U(1)^r\times K$, where $K$ is a rank $N-r-1$ semisimple Lie group, the
exact form of which depends on how the $\lm_i$ coincide with each other.
Such monopoles have $r$ topological charges.\\

{\Large{\bf III. Rational Maps}}\\

\news\ \ \ \ \ \
In this section we briefly review the recent correspondence proved by Jarvis \cite{Ja}, 
between $SU(N)$ monopoles and rational maps from the Riemann sphere into flag manifolds.
Actually the correspondence proved in \cite{Ja} is more general than this and is valid
for all compact semisimple gauge groups $G$, but in this paper we shall only be
concerned with the simplest case of $G=SU(N).$

The first step is to introduce polar coordinates, so that a point of $\R^3$ 
is given by a distance $r$ from the origin and a direction determined by a point 
$z$ on the Riemann sphere around the origin. In terms of the usual angular coordinates
$\theta,\varphi$ this is simply $z=e^{i\varphi}\tan(\theta/2).$

The Jarvis map is obtained by considering Hitchin's equation
\be
(D_r-i\Phi)s=0
\label{hitchin}
\ee
for the complex $N$-vector $s$, along each radial half-line from the origin out to infinity,
 with the direction of the half-line determined by the value of $z$. 

For the moment we shall assume that we are dealing with maximal symmetry breaking.
From the boundary conditions (\ref{cond}) we see that since at spatial infinity
$\Phi$ is in the gauge orbit of $\Phi_0$ then in the $N$-dimensional solution space
there is a one-dimensional subspace generated by the solution which decays at the
fastest rate as $r\rightarrow\infty.$ Now evaluate this solution at $r=0.$
This procedure has thus determined a line in $\C^N$ for each value of $z.$
The next step is to consider how this line varies with the direction $z$, and the
analysis shows that it varies holomorphically. The crucial ingredient here is that
the Bogomolny equation (\ref{Bog}) implies that $[D_r-i\Phi,D_\zb]=0$, so that the
operator in equation (\ref{hitchin}) commutes with the covariant derivative 
in the angular direction $D_\zb.$ It can be shown that the degree of this holomorphic map into
 \CP$^{N-1}$ is precisely the topological charge $n_1,$ and hence the map is rational.
Note that if we apply a gauge transformation then the map will be transformed
by multiplication  by a constant element of $SU(N)$, corresponding to the
gauge transformation evaluated at the origin, so that we consider only the equivalence
classes of such maps.

Now we repeat the above process but this time we consider the two-dimensional solution
space generated by the solution which decays fastest and the solution which decays the next
fastest. In the same way as above this will now define a holomorphic plane in $\C^N$
(ie. a space spanned by two holomorphic lines), which
of course will contain the holomorphic line we have already described. The degree of this
 plane is equal to the topological charge $n_2$, and so again the map is rational.
Proceeding in this way we finally arrive at the rational map 
$R:\mbox{\CP}^1 \mapsto \mbox{F}(\C^N),$ where $\mbox{F}(\C^N)$ denotes the space of total flags in $\C^N.$
This is a series of vector subspaces $0\subset V_1 \subset V_2\subset ...\subset V_{N-1}
\subset \C^N,$ where $V_i$ has dimension $i$, which is clearly the structure we have
just described.

In the above discussion the degrees refer to the elements of the homotopy group
$\pi_2(\mbox{F}(\C^N))=\Z^{N-1},$ and are given by the highest powers which
occur in some holomorphic polynomials, as described later.  
For a detailed discussion of rational maps into flag manifolds and
their relationship to monopoles the interested reader may find it useful
to consult refs. \cite{Hu3,Mu1}.

For symmetry breaking which is not maximal the picture is similar, except that now the
rational map will not be into the space of total flags, since the exponential decay of 
some of the solutions will be the same and hence some of the subspaces $V_i$ will be missing
from the flag. This of course corresponds to the fact that there will now be fewer topological
charges, and these correspond to the degrees of the maps into the vector spaces which remain
in the flag.

Because the construction of the rational map from the monopole
does not break the rotational symmetry of $\R^3$ it is a very useful approach for studying
monopoles with symmetries.
For the case of $SU(2)$ there is only one vector subspace; the space of lines in $\C^2.$
Thus the rational map is $R:\mbox{\CP}^1\mapsto\mbox{\CP}^1$, that is, a rational map
between Riemann spheres, and its degree is the sole topological charge. By explicit
construction of some symmetric maps the existence of various $SU(2)$ monopoles with
special symmetries has been proved \cite{HMS}.\\

{\Large {\bf IV. Constructing the Monopole}}
\label{construction}\\

\news\ \ \ \ \ \
The proof of the correspondence between monopoles and rational maps \cite{Ja}
involves constructing the monopole from the rational map. The starting point is
to write the Bogomolny equation (\ref{Bog}) in terms of the coordinates $r,z,\zb$
and observe that a (complex) gauge can always be chosen so that
\be
\Phi=-iA_r=-\frac{i}{2}H^{-1}\partial_r H, \ \ \
A_z=H^{-1}\partial_z H, \ \ \ A_\zb=0
\label{gauge}
\ee
where $H\in SL(N,\C)$ is a Hermitian matrix.

The Bogomolny equation is then equivalent to the single equation for $H$
\be
\pr_r\left(H^{-1}\, \pr_rH\right)+\fr{(1+|z|^2)^2}{r^2}
\pr_{\bar{z}}\left(H^{-1}\,\pr_zH\right)=0
\label{jarvis}
\ee
which we shall refer to as the Jarvis equation. Jarvis \cite{Ja} then proves that solutions
of this equation are determined by the rational map, which specifies the asymptotic boundary conditions
on $H$ for large $r.$ The analysis presented in \cite{Ja} is complicated and is not very suitable
for attempting to implement the construction explicitly, so in this section we shall present
a more explicit prescription for determining the boundary conditions on $H$ in terms of the rational
map.

For simplicity in this section we shall restrict to the case of $SU(2)$ monopoles.
 With a choice of normalization
for the Higgs field we have the boundary conditions on the monopole as
\be
\Phi=\Phi_\infty(1-\frac{n}{2r}+O(\frac{1}{r^2}))
\label{asy}
\ee
where $\Phi_\infty$ is in the gauge orbit of
$i\sigma_3=\mbox{diag}(i,-i)$, and $n$ is the topological
charge.

Any $2\times 2$ Hermitian matrix $H$, which has unit determinant, can always be
 written in the form
\be
H=\exp\bigl\{g(P-\hf)\bigr\}
\label{proj}
\ee
where $g$ is a real function and $P$ is a $2\times 2$ 
Hermitian projector, that is, $P^\dagger=P=P^2.$
A motivation for introducing projectors is that it is a useful formulation for
dealing with similar equations that arise in the context of Skyrmions \cite{IPZ}.
Examining the asymptotic boundary condition (\ref{asy}) for large $r$ we see that the
magnitude of the Higgs field at infinity is a constant and moreover the direction
of the Higgs field in the $su(2)$ algebra is independent of the radius to leading order in $1/r.$
Comparing this behaviour with equation (\ref{gauge}) for the Higgs field in terms of $H$, we
find that the leading order behaviour for large $r$ is that the profile function $g$ is independent
of the angular coordinates $z,\zb$ and the projector $P$ is a function only of the angular
coordinates. We are now going to examine the large $r$ behaviour of the solution, so we use the
above leading order result and set $g(r)$ and $P(z,\zb).$

Computing the Higgs field we obtain
\be
\Phi=-\frac{i}{2}H^{-1}\partial_r H=-\frac{i}{2}g'(P-\hf)
\label{higgs}
\ee
with magnitude
\be
\|\Phi\|^2=-\hf\mbox{tr}(\Phi^2)=\frac{g'^2}{16}=1-\frac{n}{r}+O(\frac{1}{r^2}).
\label{h2}
\ee
Integrating this equation for $g$ we obtain (there is a choice of sign here that we shall discuss
below)
\be
g=-4r+2n\log r+O(1).
\label{ag}
\ee
On substituting the form (\ref{proj}) into equation (\ref{jarvis}) and using the asymptotic
expression (\ref{ag}) we obtain the result that
\be
e^{4r}r^{-2(n+1)}(1+\vert z\vert^2)^2[PP_{z\zb}+P_\zb P_z]+O(\frac{1}{r^2})=0
\label{aeq}
\ee
where subscripts denote partial differentiation. The coefficient of the growing term
in (\ref{aeq}) must therefore vanish and we find the equation satisfied by $P$ is 
\be
(PP_z)_\zb=0.
\label{ap}
\ee
The equation $PP_z=0$ gives the instanton solutions of the \CP$^1$ $\sigma$-model in the plane
(see eg. ref \cite{Za}) and clearly these will satisfy equation (\ref{ap}). Furthermore,
as we prove in the appendix, this gives all solutions of equation (\ref{ap}).

All instanton solutions of the  \CP$^1$ $\sigma$-model are given by
\be
P=\frac{ff^\dagger}{\vert f\vert^2}
\label{ftop}
\ee
where $f$ is a 2-component column vector whose entries are holomorphic functions of $z.$
Note that the multiplication of $f$ by an overall factor does not change the projector
$P$, so that $f$ is an element of \CP$^1$.

Substituting the asymptotic behaviour (\ref{ag}) into equation (\ref{higgs}) we obtain
the expression for the Higgs field on the two-sphere at infinity 
\be
\Phi_\infty=i(2P-1).
\label{higgs0}
\ee
The topological charge, $n$, is the winding number of this map, which is equal to the
degree of the holomorphic vector $f(z)$ which is used to construct the projector via (\ref{ftop}).
Thus we conclude that the boundary condition on $H$ is determined in this simple and explicit way
in terms of the degree $n$ rational map $f(z):\mbox{\CP}^1\mapsto\mbox{\CP}^1.$

Note that (\ref{ftop}) and (\ref{higgs0}) give us an explicit expression for the Higgs field
at infinity in terms of the rational map. Naively one may think that this does not contain
very much information, since for example it is always possible to choose a (singular) 
gauge in which the
Higgs field at infinity is diagonal and constant. However, the important point is that our expression is
given in an explicit {\sl known} gauge, and therefore we have removed the gauge freedom and
are left with the physical information in the Higgs field: the fact that it is rational.

To be precise, we have not yet proved an equivalence between the rational map $f$ and the one
introduced by Jarvis \cite{Ja}. To prove this equivalence we shall now show that $f$ is 
indeed the map obtained as the scattering data.\footnote{We thank Nick Manton for suggesting this
analysis.}

In a unitary gauge there is a basis of solutions to Hitchin's equation (\ref{hitchin}) which
have the leading order large $r$ behaviour
\be
s \sim e^{-\lambda_j r}v_j
\ee
where $\lambda_j$ is an eigenvalue of $-i\Phi_\infty$ and $v_j$ is the
corresponding eigenvector. In the $SU(2)$ case, when $\lambda_1=-\lambda_2=1$,
the scattering map is determined by the decaying solution or more fundamentally
by the solution associated with the $\lambda_1=1$ eigenspace.
Recall that the scattering map is obtained by evaluating this solution at the
origin $r=0.$ Now, in the gauge (\ref{gauge}) Hitchin's equation is trivialised
to $\partial_r s=0$, so the solutions are $r$ independent and hence the scattering
map is the eigenvector of $-i\Phi_\infty$ with eigenvalue one. Thus all that remains to
be shown is that $f$ is the eigenvector of $-i\Phi_\infty$ with eigenvalue one.
Using the explicit expression (\ref{higgs0}) and the definition of the projector
(\ref{ftop}) this is elementary as
\be
-i\Phi_\infty f=(2P-1)f=(\frac{2ff^\dagger}{\vert f\vert^2}-1)f=f.
\ee

On a minor point, it is worth while making a comment about the choice of sign made in equation (\ref{ag})
when taking the square root and integrating equation (\ref{h2}). If the opposite choice of sign is made
then following through the analysis we find that the boundary condition is determined by an
anti-holomorphic map. Thus this choice of sign is merely an orientation and determines whether we
wish monopoles to correspond to holomorphic or anti-holomorphic rational maps.

The construction of a monopole from its rational map is now clear. Choose a rational map $f(z)$
and calculate the associated projector (\ref{ftop}). Then compute the solution of the 
Jarvis equation (\ref{jarvis}) satisfying the boundary condition that for large $r$
\be
H\sim \exp(r(2-4P)).
\ee

Obviously this construction is not easy to implement explicitly in practice, since it still requires
the solution of a nonlinear partial differential equation. In this sense it is not as powerful as
say the ADHMN construction \cite{Na}, which reduces the problem to solving a set of nonlinear
matrix ordinary differential equations plus a further linear system of ordinary differential equations.
The advantage is that for the construction discussed here the data is free, in that any rational map
is allowed, whereas in the ADHMN construction the Nahm data must satisfy complicated constraints
(including the aforementioned set of nonlinear ordinary differential equations). Thus even using
the ADHMN construction very few explicit examples of monopole solutions are known. There is always
an inherent difficulty associated with solving the monopole equations and the difference between
these two alternative constructions is whether the main difficulty resides in performing the construction
or specifying the data upon which the construction is performed.

There are simplifying special cases for which we are able to perform the construction explicitly,
the easiest example being the rational map $f=(1,z)^t$, which corresponds to the spherically symmetric
$SU(2)$ 1-monopole. In this case the asymptotic behaviour, $g(r)$ and $P(z,\zb)$,
 is valid for all $r$ and
substituting (\ref{proj}) into the Jarvis equation gives the following ordinary differential
equation for the profile function
\be
g''+\frac{2}{r^2}(1-e^g)=0.
\label{g1}
\ee
The large $r$ boundary condition  $g\sim -4r$, together with the
condition $g(0)=0$, which is required for $H$ to be well-defined at the origin,
determines the unique solution of (\ref{g1}) as
\be 
g=2\log(2r/\mbox{sinh}2r).
\label{g1s}
\ee
This gives  the well-known 1-monopole solution and comparing the asymptotic expansion
of (\ref{g1s}) with equation (\ref{ag}) we verify that $n=1$, so we see explicitly
that the topological charge is determined as the degree of the rational map and there
is no freedom in the profile function once the map has been
specified.

In the following section we provide some explicit examples of solutions to the Jarvis
equation, corresponding to spherically symmetric $SU(N)$ monopoles with various
symmetry breakings. We present the rational maps and describe how the solutions of
the Jarvis equation are obtained from these in terms of harmonic maps into
\CP$^{N-1}$.\\

{\Large {\bf V. Harmonic Maps and Spherical Monopoles}}
\label{sec-harm}\\

\news\ \ \ \ \ \
In the first part of this section we briefly review some facts that we shall need 
about harmonic maps of the \CP$^{N-1}$ $\sigma$-model in the plane. These results can be
found in, for example, ref. \cite{Za}.\\

{\Large {\bf A. Harmonic Maps}}
\label{hmaps}\\

The harmonic map (or $\sigma$-model) equations for the \CP$^{N-1}$ model are given by
\be
[P_{z\zb},P]=0
\label{hmap}
\ee
where $P$ is an $N\times N$ Hermitian projector.

As stated earlier, one set of solutions to these equations are the instantons given by
\be
P(f)=\frac{ff^\dagger}{\vert f\vert^2}
\label{ftop2}
\ee
where $f(z)$ is an $N$-component column vector which is a holomorphic function of $z$ and
 whose degree is equal to the topological
charge of the $\sigma$-model. Another set of solutions are the anti-instantons,
which have the same form but this time $f$ is an anti-holomorphic function, and then
the $\sigma$-model topological charge is minus the degree of $f.$

For $N=2$ these are all the finite action solutions, but for $N>2$ there are other non-instanton
solutions. These can be described by introducing the operator $\pp$ defined  by its action   
on any vector $f\in \C^N$ as
\be
\pp f=\pr_z f- \fr{f \,(f^\dg \,\pr_z f)}{|f|^2}
\ee
and then define further vectors $\pp^k f$ by induction:
$\pp^k f=\pp(\pp^{k-1} f)$.

To proceed further we note the following useful properties of
$\pp^k f$ when $f$ is holomorphic:
\begin{eqnarray}
\label{bbb}
&&(\pp^k f)^\dg \,\pp^l f=0, \hs \hs \hs k\neq l\acc
&&\pr_{\bar{z}}\left(\pp^k f\right)=-\pp^{k-1} f \fr{|\pp^k
f|^2}{|\pp^{k-1} f|^2},
\hs \hs
\pr_{z}\left(\fr{\pp^{k-1} f}{|\pp^{k-1} f|^2}\right)=\fr{\pp^k
f}{|\pp^{k-1}f|^2}.
\label{aaa}
\end{eqnarray}
These properties either follow directly from the definition of $\pp$
or are easy to prove \cite{Za}.
It is also convenient to define projectors $P_k$ corresponding to the
family of vectors $\pp^k f$  as 
\be
P_k=P(\pp^k f), \ \ k=0,..,N-1.
\ee
Applying $\pp$ a total of $N-1$ times to a holomorphic vector gives an anti-holomorphic
vector, so that a further application of $\pp$ gives the zero vector
and hence no corresponding projector. 

The projectors $P_k$ are solutions of the harmonic map equations (\ref{hmap})
and all solutions can be found in this way by starting with an appropriate
holomorphic vector $f$. In the \CP$^1$ case the operator $\pp$
converts a holomorphic vector to an anti-holomorphic vector, that is, instantons
to anti-instantons and these are all the solutions in this case.

Note that the projectors obtained from this sequence always satisfy
the relation $\sum_{k=0}^{N-1}P_k=1$.

For connecting harmonic maps with monopoles it is useful to recall the following
interpretation of the non-instanton solutions \cite{Za}. From a holomorphic
vector $f$ form the exterior product of $f$ and its derivatives as
\be
h^k=f\wedge\partial_z f\wedge ..\wedge \partial_z^k f, \ \ \ k=0,...,N-1.
\label{defh}
\ee
Thus $h^k$ is holomorphic, though it is an element of a larger dimensional space;
it may be represented as a totally anti-symmetric tensor with $k+1$ indices.
With this notation it may then be shown that
\be
\bar h^{k-1}\cdot h^k \simeq \pp^k f
\label{ip}
\ee
where $\cdot$ denotes the summation over all the indices of $h^{k-1}$ and all but the first
index of $h^{k}$.
Here $\simeq$ denotes that two vectors are equal up to an overall factor,
which is the important equivalence since we are dealing with elements of projective
spaces.
 Equation (\ref{ip})  leads to the relation
\be
\mbox{deg}(\pp^k f)=\mbox{deg}(h^k)-\mbox{deg}(h^{k-1})
\label{degrees}
\ee
where the left-hand side is defined as the $\sigma$-model topological charge of the
projector $P_k=P(\pp^k f),$ and $\mbox{deg}(h^k)$ is the highest power of $z$ which
occurs in the holomorphic tensor $h^k.$ Thus the non-instanton solutions may be interpreted
as special mixtures of instantons and anti-instantons.\\

{\Large {\bf B. Spherical Monopoles}}\\

In section IV we saw that the rational map for the spherically symmetric
$SU(2)$
1-monopole is given by $f(z)=(1,z)^t.$ This map is spherically symmetric in the sense that
a rotation in $\R^3$, which is realised as an $SU(2)$ M\"obius transformation
\be
z\mapsto\widetilde z=\frac{\alpha z+\beta}{-\bar\beta z+\bar\alpha}, \ \ \mbox{with} 
\ \ \vert\alpha\vert^2+\vert\beta\vert^2=1
\ee
can be compensated by a constant $SU(2)$ gauge transformation.
Explicitly
\be
f(\widetilde z)\simeq\pmatrix{\bar\alpha&-\bar\beta\cr \beta & \alpha\cr}f(z).
\label{ss1}
\ee
This is the only spherically symmetric map into \CP$^1$ which has positive degree and hence there
are no more spherically symmetric $SU(2)$ monopoles. For $SU(N)$ we first require
spherically symmetric maps into \CP$^{N-1}$ and these are given by
\be
f=(f_0,...,f_j,...,f_{N-1})^t, \ \ \mbox{where} \ \ f_j=z^j\sqrt{{N-1}\choose j}
\label{smap}
\ee
and ${N-1}\choose j$ denote the binomial coefficients. It can be shown that these
maps are spherically symmetric by an explicit presentation of the compensating
transformation as in (\ref{ss1}). Of course there are other spherically symmetric maps
which are obtained by embedding the above maps and setting all other entries to be zero.

For a spherically symmetric $SU(N)$ monopole we require a 
 rational map into the space of total flags $\mbox{F}(\C^N),$ which has
 spherical symmetry. Thus we need an explicit representation of the
holomorphic line, the holomorphic plane (which contains the line),...etc.
As we shall see in more detail below, we take each $k$-dimensional subspace to be the
 space spanned by the vectors $f,\partial_z f,...,\partial_z^{k-1}f,$ where $f$ is the
spherical map (\ref{smap}). Note that these
are precisely the spaces $h^{k-1}$ defined in (\ref{defh}). Thus the topological charges
of the monopole, $n_k$, are given by 
\be
n_k=\mbox{deg}(h^{k-1}), \ \ k=1,...,N-1.
\ee
Hence from (\ref{degrees}) it is clear that the monopole topological charges are therefore not
equal to the $\sigma$-model topological charges of the harmonic maps from which we shall create them.
The exception to this statement is the case $N=2$, where all the harmonic maps are instantons
 and then the only degree is $\mbox{deg}(h^0)$ which in this case is equal to the $\sigma$-model
topological charge.

The degree of the map (\ref{smap}) is $N-1$ and hence it is easy to calculate the degree of $h^k$
from (\ref{defh}) which, after taking into account the anti-symmetry, gives
\be
n_k=\mbox{deg}(h^{k-1})=k(N-k), \ \ \ \ \ k=1,...,N-1.
\label{charges}
\ee
Thus we have computed the monopole topological charges and now it remains to construct
the corresponding solution of the Jarvis equation. The $SU(N)$ generalization of the $SU(2)$
form given in (\ref{proj}) is to take a sum of the $N-1$ projectors 
\be
H=\exp\bigl\{g_0(P_0-\frac{1}{N})+g_1(P_1-\frac{1}{N})+...+g_{N-2}(P_{N-2}-\frac{1}{N})\bigr\}
\label{projsun}
\ee
where $g_k(r)$ for $k=0,...,N-2,$ are profile functions. Recall that the projector $P_{N-1}$
is a linear combination of the other projectors plus the identity matrix, which is why it is 
not included in the above formula. The profile functions satisfy the regularity condition
$g_k(0)=0$, and have a linear growth in $r$ for large $r$, the coefficients of which determine
the symmetry breaking pattern. Once the symmetry breaking is specified the profile functions
are, of course, uniquely determined; since there is a one-to-one correspondence between
monopoles and rational maps. We shall illustrate this explicitly in the following with some 
examples.\\

{\Large{\bf C. SU(3) Examples}}
\label{sec-su3examples}\\

For $N=3$, with symmetry breaking to $U(1)\times U(1)$, the charges (\ref{charges}) 
are $(n_1,n_2)=(2,2).$ From (\ref{smap}) the holomorphic 
line is given by $f=(1,\sqrt{2}z,z^2)^t$ and the plane is spanned by $f$ and $f_z.$ 
The $SU(3)$ case has a simplifying feature, in that the holomorphic plane in $\C^3$ can
be specified by giving a line orthogonal to the plane; which will then be anti-holomorphic.
This line is given by
\be
f_\perp =\overline{f \times f_z} =\sqrt{2}(\zb^2,-\sqrt{2}\zb,1)^t
\label{perp}
\ee
which is clearly anti-holomorphic and by construction is orthogonal to the holomorphic plane,
 that is, $f_\perp^\dagger f= f_\perp^\dagger f_z=0.$ By inspection of (\ref{perp}) the
plane has degree two and clearly has spherical symmetry (compare the structure of 
$f_\perp$ and $f$). Hence in this case it is simple to see that the charge is $(2,2)$.
However, as an illustration of the general formalism we shall also present this example in terms
of the notation described above. Thus we find 
\be
h^0=\pmatrix{1\cr\sqrt{2}z\cr z^2\cr}, \ \ \ 
h^1=\pmatrix{0 & \sqrt{2} & 2z\cr
-\sqrt{2} & 0 &\sqrt{2}z^2 \cr
-2z & -\sqrt{2}z^2 & 0}
\label{h22}
\ee
giving $(n_1,n_2)=(\mbox{deg}(h^0),\mbox{deg}(h^1))=(2,2)$.

Taking the $h^k$ from (\ref{h22}) we construct the associated projectors, using (\ref{ip}),
and insert these into the form for $H$ given in (\ref{projsun}). This gives a solution of the
Jarvis equation provided the profile functions satisfy the ordinary differential equations
\begin{eqnarray}
-g_{0}''+\fr{2}{r^2}\left(e^{g_0-g_1}-1\right)+\fr{2}{r^2}\left(e^{g_1}
-1\right)&=&0\nonumber\\
-g_{1}''-\fr{2}{r^2}\left(e^{g_0-g_1}-1\right)+\fr{4}{r^2}\left(e^{g_1}
-1\right)&=&0.
\label{su3}
\end{eqnarray}
The Higgs field is given in terms of the solution of the Jarvis equation by (\ref{gauge})
and the eigenvalues and topological charges can simply be read off by restricting to a given radial line, 
say $z=0,$ which gives
$\Phi=\fr{i}{6}\mbox{diag}(g_{1}'-2g_{0}',\,g_{0}'-2g_{1}',\,g_{0}'+g_{1}').$

Each profile function has an asymptotic expansion of the form
\be
g_k=-\alpha_k r+\beta_k\log r+\log\gamma_k+O(\frac{1}{r})
\ee
with the $\alpha_k$ determined by the vacuum expectation value of the Higgs field.
Comparing with (\ref{cond}) for this case we have that
\be
\lambda_1=\frac{2\alpha_0-\alpha_1}{6}, \ 
\lambda_2=\frac{2\alpha_1-\alpha_0}{6}, \
\lambda_3=-\frac{\alpha_0+\alpha_1}{6}, \
n_1=\frac{2\beta_0-\beta_1}{3}, \
n_2=\frac{\beta_0+\beta_1}{3}.
\label{chsu3}
\ee
It is simple to verify that the topological charge is $(2,2)$ without resorting to
an explicit solution of the profile function equations (\ref{su3}). Maximal symmetry
breaking implies that $\alpha_0>\alpha_1>0$, so that the terms in (\ref{su3}) which
contain exponentials of profile functions do not contribute to the leading order behaviour
which is $O(\frac{1}{r^2})$. The coefficients of this leading order term then simply give
that $\beta_0=4,\ \beta_1=2$, which when substituted into (\ref{chsu3}) confirms that
$(n_1,n_2)=(2,2).$

The explicit solutions for the profile functions can be obtained, for example, if we choose
$\Phi_0=\mbox{diag}(2,0,-2)$, then the solution is $g_0=2g_1=2g$, where $g$ is the 
1-monopole profile function defined in (\ref{g1s}).

If we now consider the case of minimal symmetry breaking then the topological
charges which survive will be unchanged, but the magnetic weights will not
be given by the topological charges which do not survive. As an example
consider the symmetry breaking to $U(1)\times SU(2)$ given by 
$\Phi_0=\mbox{diag}(1,-\frac{1}{2},-\frac{1}{2}).$
From (\ref{chsu3}) this corresponds to setting $\alpha_0=3, \ \alpha_1=0.$
The previous analysis of the profile function equations must now be modified
to take into account the fact that exponentials of profile functions may
now contribute to leading order (this happens whenever any of the $\alpha_k$ 
coincide or are zero, and corresponds to changing the symmetry breaking pattern).
In this case it is easy to see that (\ref{su3}) requires that 
$\beta_0=3,\ \beta_1=0, \ \gamma_1=\frac{1}{2}$ which gives the values
$(n_1,[n_2])=(2,[1])$, where we have used the notation that square brackets
denote magnetic weights rather than topological charges. 

The profile function equations that we obtain are related to those derived from
the ansatz based approach of Bais et al \cite{BWe,BW} and the 
methods employed there can be adapted to solve for the profile functions 
explicitly. This method requires a careful limiting procedure to be taken
to deal with non-maximal symmetry breaking. In section VI 
we shall see that the solutions for monopoles in hyperbolic space are
obtained without the need for this limiting procedure and the Euclidean case can
then be obtained from the natural limit in which the curvature of hyperbolic space
tends to zero.

For this example the solution is (see section VI)
\bea
g_0&=&\log\fr{81\,r^4}{4\,[(-3r-1)e^{-r}+e^{2r}]\,[(3r-1)e^r+e^{-2r}]} \nonumber\\
g_1&=&\log\fr{9\,r^2\,[(-3r-1)e^{-r}+e^{2r}]}{2\,[(3r-1)e^r+e^{-2r}]^2}
\label{su3soln}
\eea
and it can be checked that the asymptotic properties are as stated above.

Spherically symmetric monopoles of lower charge, such as the $(1,1)$ monopole,
can be obtained in a similar way by embedding the spherically symmetric maps
(\ref{smap}) of lower degree.\\

{\Large {\bf D. SU(4) Examples}}\\

For maximally broken $SU(4)$ the charge, from (\ref{charges}), is $(3,4,3)$
and the associated profile function equations are
\begin{eqnarray}
-g_{0}''+\fr{3}{r^2}\left(e^{g_0-g_1}-1\right)
+\fr{3}{r^2}\left(e^{g_2}-1\right)&=&0\nonumber\\
-g_{1}''-\fr{3}{r^2}\left(e^{g_0-g_1}-1\right)+\fr{4}{r^2}\left(e^{g_1
-g_2}-1\right)+\fr{3}{r^2}\left(e^{g_2}-1\right)&=&0\nonumber\\
-g_{2}''-\fr{4}{r^2}\left(e^{g_1-g_2}-1\right)+
\fr{6}{r^2}\left(e^{g_2}-1\right)&=&0.
\label{SU4eq}
\end{eqnarray}
As for the $SU(3)$ case it is a simple task to confirm the topological charge
by a leading order analysis of this set of equations. For the choice 
$\Phi_0=\mbox{diag}(3,1,-1,-3)$, corresponding to equal monopole masses, the explicit
solution is $g_0/3=g_1/2=g_2=g$, where $g$ is given by (\ref{g1s}).

There are several possible symmetry breakings and in each case it is a simple matter
to determine both the topological charges and magnetic weights by an analysis
of equations (\ref{SU4eq}).

For $\Phi_0=\mbox{diag}(1,\fr{1}{2}, \fr{1}{2},-2)$ the symmetry breaking is
$U(1)\times SU(2)\times U(1)$ and the charge is $(3,[3],3).$ The corresponding explicit
solution is 
\begin{eqnarray}
g_0&=&\log\fr{625\,r^6}{\left[-25e^{-2r}-(30r-24)e^{-r}+e^{4r}\right]
\left[25e^{2r}-(30r+24)e^r-e^{-4r}\right]}\nonumber\acc
g_1&=&\log\fr{25\,r^4\,[25e^{2r}-(30r+24)e^r-e^{-4r}]}
{2\left[-25e^{-2r}-(30r-24)e^{-r}+e^{4r}\right]\left[6e^{-2r}
+(5r-6)e^{3r}-(5r+6)e^{-3r}+6e^{2r}\right]}\nonumber\acc
g_2&=&\log\fr{50\,r^2\left[6e^{-2r}+(5r-6)e^{3r}-(5r+6)e^{-3r}+6e^{2r}\right]}
{\left[-25e^{-2r}-(30r-24)e^{-r}+e^{4r}\right]^2}.
\end{eqnarray}

Choosing $\Phi_0=\mbox{diag}
(\fr{3}{4},\fr{3}{4},-\fr{1}{4},-\fr{5}{4})$ gives the symmetry breaking 
$SU(2)\times U(1)\times U(1)$ with charge $([2],4,3)$  and solution
\begin{eqnarray}
g_0&=&\log \fr{256\,r^6}{9\,\left[(4r+3)e^{-3r/2}+e^{5r/2}-4e^{r/2}\right]
\left[(4r-3)e^{3r/2}-e^{-5r/2}+4e^{-r/2}\right]}\nonumber\acc
g_1&=&\log \fr{16\,r^4\,\left[(4r-3)e^{3r/2}-e^{-5r/2}+4e^{-r/2}\right]}{
3\,\left[(4r+3)e^{-3r/2}+e^{5r/2}-4e^{r/2}\right]\left[(-4r-1)e^r+(4r-1)e^{-r}
+e^{-3r}+e^{3r}\right]}\nonumber\acc
g_2&=&\log
\fr{16\,r^2\,\left[(-4r-1)e^r+(4r-1)e^{-r}+e^{-3r}+e^{3r}\right]}{
3\,\left[(4r+3)e^{-3r/2}+e^{5r/2}-4e^{r/2}\right]^2}.
\end{eqnarray}

By taking two pairs of eigenvalues to be equal, for example 
$\Phi_0=\mbox{diag}(\fr{1}{2},\fr{1}{2},-\fr{1}{2},-\fr{1}{2})$, the
symmetry is broken to $SU(2)\times U(1)\times SU(2)$. In this case the
charge is $([2],4,[2])$ and the
profile functions are given by
\be
g_0=\log \fr{r^6}{9\,(r\cosh r-\sinh r)^2},\hs
g_1=\log \fr{r^4}{3\,(\sinh^2 r-r^2)},\hs
g_2=\log \fr{r^2 \,(\sinh^2 r-r^2)}{3\,(r\cosh r-\sinh r)^2}.
\ee

Finally, minimal symmetry breaking to $U(1)\times SU(3)$ occurs when
three eigenvalues coincide, say $\Phi_0=\mbox{diag}(3,-1,-1,-1)$ and
this gives a charge $(3,[2],[1])$ with solution
\begin{eqnarray}
g_0&=&\log
\fr{1024\,r^6}{9\,[e^r(8r^2-4r+1)-e^{-3r}]\,[e^{3r}-e^{-r}(8r^2+4r+1)]}\nonumber\acc
g_1&=&\log \fr{16\,r^4\,
[e^{3r}-e^{-r}(8r^2+4r+1)]}{3\,[e^r(8r^2-4r+1)-e^{-3r}]
\,[e^{2r}(2r^2-r)+e^{-2r}(2r^2+r)]}\nonumber\acc
g_2&=&\log\fr{64\,r^2\, [e^{2r}(2r^2-r)+e^{-2r}(2r^2+r)]}
{3\,[e^r(8r^2-4r+1)-e^{-3r}]^2}.
\label{minsu4}
\end{eqnarray}\\

{\Large {\bf VI. Hyperbolic Monopoles}}
\label{sec-hyperbolic}\\

\news\ \ \ \ \ \
Hyperbolic monopoles are solutions of the Bogomolny equation (\ref{Bog}) in which
Euclidean space $\R^3$ is replaced by hyperbolic 3-space, which we denote by $\HH^3_\kk$, where
$-\kk^2$ is the curvature of hyperbolic space. They were first studied by Atiyah
\cite{At2}, who observed that $S^1$ invariant instantons can be interpreted as hyperbolic
monopoles. Often hyperbolic monopoles turn out to be easier to study than the
Euclidean case and we shall see this explicitly in the following. It has long been
expected that in the limit as the curvature of hyperbolic space tends to zero then
Euclidean monopoles are recovered, but only recently has this been rigorously established
\cite{JN}. In this section we shall adapt the methods of section V to
the hyperbolic case to obtain spherically symmetric $SU(N)$ monopoles. The $SU(2)$ 1-monopole
solution has been obtained before \cite{Cha}, as a circle invariant instanton, but
we believe that all our multi-monopole solutions are new. By explicitly taking the zero
curvature limit we recover the Euclidean monopole solutions and explain why 
this is a simpler way to obtain these solutions than to consider the Euclidean case
from the beginning.

Perhaps the most familiar description of $\HH^3_\kk$ is as the interior of the unit ball.
In terms of angular coordinates $z,\zb$ and radial coordinate $\rho\in[0,1)$ the metric is
\be
ds^2=\frac{4}{\kk^2(1-\rho^2)^2}(d\rho^2+\rho^2 \frac{4dzd\zb}{(1+\vert z\vert^2)^2})
=dr^2+\frac{\mbox{sinh}^2(\kk r)}{\kk^2} \frac{4dzd\zb}{(1+\vert z\vert^2)^2}
\ee
where we have introduced $r$, the hyperbolic distance from the origin, through the
relation
\hbox{$\rho=\mbox{tanh}(\kk r/2).$}

The radial scattering analysis proceeds as in the Euclidean case 
with the upshot that the Jarvis equation (\ref{jarvis}) in hyperbolic space
 becomes \cite{JN,Ja}
\be
\pr_r\left(H^{-1}\, \pr_rH\right)+\fr{\kk^2(1+|z|^2)^2}{\mbox{sinh}^2 (\kk r)}
\pr_{\bar{z}}\left(H^{-1}\,\pr_zH\right)=0.
\label{hypjarvis}
\ee
Note that in the zero curvature limit, $\kk\rightarrow 0$, the Euclidean
equation (\ref{jarvis}) is recovered.

Solutions of (\ref{hypjarvis}) can be obtained using the form (\ref{projsun}), with
the same harmonic maps, but leading to modified equations for the profile functions.
The equations for the monopole fields in terms of $H$ are still given by (\ref{gauge}),
but with $r$ now being hyperbolic distance. Hence the asymptotic boundary conditions
remain the same as in the Euclidean case and together with the requirement 
that the profile functions vanish at the origin this determines a unique solution
for any given choice of vacuum expectation value $\Phi_0$.

For the $SU(2)$ 1-monopole there is just one profile function, which must satisfy
the equation 
\be
g''+\frac{2\kk^2}{\mbox{sinh}^2 (\kk r)}(1-e^g)=0.
\label{hypg1}
\ee
If we again normalize the Higgs field to have unit magnitude then
the boundary conditions on the profile function are $g(0)=0$ and  
$g(r)\sim -4r$ for large $r.$ The solution is 
\be
g=2\log\frac{(2+\kk)\mbox{sinh} (\kk r)}{\kk \mbox{sinh}((2+ \kk) r)}
\label{hypg}
\ee
which gives the known $SU(2)$ hyperbolic 1-monopole \cite{Cha}.\\

{\Large{\bf A. SU(3) Examples}}\\

The profile function equations for the $SU(3)$ charge $(2,2)$ hyperbolic monopole are
\begin{eqnarray}
-g_{0}''\frac{\mbox{sinh}^2 (\kk r)}{\kk^2}+2\left(e^{g_0-g_1}-1\right)+2\left(e^{g_1}
-1\right)&=&0\nonumber\\
-g_{1}''\frac{\mbox{sinh}^2 (\kk r)}{\kk^2}-2\left(e^{g_0-g_1}-1\right)+4\left(e^{g_1}
-1\right)&=&0.
\label{hypsu3}
\end{eqnarray}
For equal monopole masses, with $\Phi_0=\mbox{diag}(2,0,-2)$, the solution is
$g_0=2g_1=2g$, with $g$ given by (\ref{hypg}). For general $\Phi_0$, including minimal
symmetry breaking, we now describe how the solution to (\ref{hypsu3}) can be obtained
using Hirota's method. 

Introducing the tau-functions $\tau_0,\tau_1$ via the transformation
\be
g_0=\log\frac{\mbox{sinh}^4 (\kk r)}{\tau_0\tau_1\kk^4}, \ \ \
g_1=\log\frac{\tau_0\mbox{sinh}^2 (\kk r)}{\tau_1^2\kk^2}
\label{tausu3}
\ee
converts equation (\ref{hypsu3}) into Hirota bilinear form
\be
{\cal D}^2\tau_i . \tau_i+4\tau_{i+1}\tau_{i-1}=0, \ \ \ \ \ \ \ i=0,1
\label{bisu3}
\ee
where we have defined $\tau_{-1}=\tau_{2}= 1$, and ${\cal D}$ is
the Hirota derivative defined by \cite{Hi}
\be
{\cal D}^m \alpha . \beta = (\partial_r-\partial_{\widetilde r})^m 
\alpha(r)\beta(\widetilde r)\vert_{\widetilde r=r}.
\label{hirota}
\ee
The Hirota derivative has many special properties which make the construction of
solutions to bilinear equations such as (\ref{bisu3}) an elegant procedure.
In particular from (\ref{hirota}) it is clear that its action on exponential functions 
takes the simple form
\be
{\cal D}^m e^{\alpha_1 r} . e^{\alpha_2 r} = (\alpha_1-\alpha_2)^m e^{(\alpha_1+\alpha_2)r}
\label{prop}
\ee
Using this property, together with the bilinear form of the equation, means that it
is a simple task to find solutions which are finite sums of exponential functions; 
in the context of integrable soliton equations, such as the KdV equation, solutions
of bilinear equations which are a finite sum of exponentials correspond to multi-solitons
\cite{Hi}.

Note that the bilinear equations (\ref{bisu3}) are independent of the curvature, $-\kk^2$, 
so in particular these equations are the ones which also arise for Euclidean monopoles.
However, the transformation (\ref{tausu3}) involves $\kk$ which means that the boundary
conditions on the tau-functions are $\kk$ dependent, and this is of crucial importance.
For hyperbolic monopoles, ie. $\kk\neq 0$, the solutions satisfying the required boundary
conditions are always given as a simple sum of exponentials, whereas for Euclidean monopoles
the boundary conditions (except for the case of maximal symmetry breaking) mean that
 the solutions are not so simple and involve a sum of products of exponentials and polynomials.
By taking the limit $\kk \rightarrow 0$ of a hyperbolic solution the more complicated
Euclidean solutions are obtained and this is perhaps the most natural method to construct
 Euclidean monopoles.

As an example, consider the minimal symmetry breaking of $SU(3)$ obtained from
$\Phi_0=\mbox{diag}(1,-\frac{1}{2},-\frac{1}{2}).$ As discussed in section
V this choice of $\Phi_0$ corresponds to the large $r$ boundary
conditions $g_i\sim {-\alpha_i r}$, with $\alpha_0=3$, $\alpha_1=0.$
Comparing this with the transformation (\ref{tausu3}) gives the large $r$ boundary conditions
\be
\tau_0\sim A_0 e^{(2+2\kk)r}, \ \ \ \tau_1\sim A_1 e^{(1+2\kk)r}
\label{lo}
\ee
for some constants $A_i.$ The requirement that $g_i(0)=0$ gives the conditions at the
origin that $\tau_0=r^2+...$, $\tau_1=r^2+...$ as $r\rightarrow 0.$
Using the leading order behaviour (\ref{lo}) together with the properties of
the Hirota derivative it is a simple task to find the following explicit solution
\bea
\tau_0&=&
\frac{2\kk e^{(2+2\kk)r}-(3+4\kk)e^{-r}+
(3+2\kk)e^{-(1+2\kk)r}}{(3+2\kk)(3+4\kk)\kk}
\nonumber\\
\tau_1&=&
\frac{(3+2\kk)e^{(1+2\kk)r}-(3+4\kk)e^{r}+2\kk e^{-(2+2\kk)r}}{(3+2\kk)(3+4\kk)\kk}.
\eea
As claimed above, we see that there is only an exponential dependence on $r$; this corresponds
to the fact that hyperbolic monopoles approach the vacuum value exponentially, rather than
algebraically like Euclidean monopoles. 

Taking the limit $\kk\rightarrow 0$ this solution becomes
\be
\tau_0=\frac{2}{9}(e^{2r}-(3r+1)e^{-r}), \ \ \ \
\tau_1=\frac{2}{9}((3r-1)e^{r}+e^{-2r})
\label{su3limit}
\ee
so we see the emergence of the algebraic factors.
Substituting (\ref{su3limit}) into (\ref{tausu3}) we obtain the Euclidean monopole
solution given by (\ref{su3soln}).\\

{\Large{\bf B. SU(4) Examples}}\\

The $SU(4)$ equations are 
\begin{eqnarray}
-g_{0}''\frac{\mbox{sinh}^2 (\kk r)}{\kk^2}+3\left(e^{g_0-g_1}-1\right)
+3\left(e^{g_2}-1\right)&=&0\nonumber\\
-g_{1}''\frac{\mbox{sinh}^2 (\kk r)}{\kk^2}-3\left(e^{g_0-g_1}-1\right)+4\left(e^{g_1
-g_2}-1\right)+3\left(e^{g_2}-1\right)&=&0\nonumber\\
-g_{2}''\frac{\mbox{sinh}^2 (\kk r)}{\kk^2}-4\left(e^{g_1-g_2}-1\right)+
6\left(e^{g_2}-1\right)&=&0.
\label{hypSU4eq}
\end{eqnarray}
The solution $g_0/3=g_1/2=g_2=g$, with $g$ given by (\ref{hypg}), corresponds
to maximal symmetry breaking with $\Phi_0=\mbox{diag}(3,1,-1,-3).$

To obtain the solution for arbitrary $\Phi_0$ we introduce the tau-functions as
\be
g_0=\log\frac{\mbox{sinh}^6 (\kk r)}{\tau_0\tau_2\kk^6}, \ \
g_1=\log\frac{\tau_0\mbox{sinh}^4 (\kk r)}{\tau_1\tau_2\kk^4}, \ \
g_2=\log\frac{\tau_1\mbox{sinh}^2 (\kk r)}{\tau_2^2\kk^2} 
\label{tausu4}
\ee
which transforms the equation into the Hirota form
\be
{\cal D}^2\tau_i . \tau_i+2(1+i)(3-i)\tau_{i+1}\tau_{i-1}=0, \ \ \ \ \ \ \ i=0,1,2
\label{bisu4}
\ee
where $\tau_{-1}=\tau_3=1$.

As an example we give the solution for minimal symmetry breaking with 
$\Phi_0=\mbox{diag}(3,-1,-1,-1)$, which is
\bea
\tau_0&=&3\frac{\kk^2e^{(3+3\kk)r}-(3\kk^2+5\kk+2)e^{(-1+\kk)r}+(3\kk^2+8\kk+4)e^{(-1-\kk)r}
-(\kk^2+3\kk+2)e^{(-1-3\kk)r}}{8(1+\kk)(2+\kk)(2+3\kk)\kk^2}\nonumber\acc
\tau_1&=&3\frac{(2+\kk)\mbox{cosh}((2+4\kk)r)-(4+4\kk)\mbox{cosh}((2+2\kk)r)+(2+3\kk)\mbox{cosh}(
2r)}{8(1+\kk)(2+\kk)(2+3\kk)\kk^2}\nonumber\acc
\tau_2&=&3\frac{(\kk^2+3\kk+2)e^{(1+3\kk)r}-(3\kk^2+8\kk+4)e^{(1+\kk)r} 
+(3\kk^2+5\kk+2)e^{(1-\kk)r}-\kk^2e^{-(3+3\kk)r}}{8(1+\kk)(2+\kk)(2+3\kk)\kk^2}.\nonumber
\eea
Taking the zero curvature limit results in
\bea
\tau_0&=&\frac{3}{32}(e^{3r}-(8r^2+4r+1)e^{-r})\nonumber\\
\tau_1&=&\frac{3r}{16}((2r-1)e^{2r}+(2r+1)e^{-2r})\nonumber\\
\tau_2&=&\frac{3}{32}((8r^2-4r+1)e^{r}-e^{-3r})
\eea
which is the Euclidean monopole solution given in (\ref{minsu4}).\\

{\Large {\bf VII. Conclusion}}\\

\news\ \ \ \ \ \
We have studied in some detail the construction of $SU(N)$ monopoles from scattering data
which consists of a rational map of the Riemann sphere into a flag manifold.
Explicit solutions have been obtained in the case of spherical symmetry and we have
shown how these solutions involve harmonic maps of the plane into \CP$^{N-1}$.
This approach was generalized to the case of hyperbolic monopoles and new spherically
symmetric solutions found, whose zero curvature limit was investigated explicitly.

The Jarvis equation is integrable, but in this paper we have made no use of
the Lax pair. The precise description of the boundary
conditions in terms of the rational map makes this a very convenient formulation
of the Bogomolny equation and it may prove useful to undertake a classical 
inverse scattering study. An alternative, which is currently under investigation, is the numerical solution
 of the Jarvis equation, which is more promising than a numerical solution of the Bogomolny 
equation since the boundary conditions can be specified in a simple manner to ensure
the existence of a unique solution.\\

\section*{Acknowledgements}
\news\ \ \ \ \ \
Many thanks to Conor Houghton, Stuart Jarvis, Nick Manton, Michael Singer and Wojtek Zakrzewski for
useful discussions. PMS acknowledges the EPSRC for an Advanced Fellowship
and the grant GR/L88320.\\

\appendix
\setcounter{equation}{0} 
\section*{Appendix}
\news\ \ \ \ \ \
\renewcommand{\theequation}{A.\arabic{equation}}
Let $P(z,\zb)$ be a $2\times 2$ Hermitian projector. In this appendix we
prove that the only solutions of the equation
\be
(PP_z)_\zb=0
\label{oureqn}
\ee
are the $\sigma$-model instantons given by
\be
PP_z=0.
\label{sigma}
\ee
Let $F=PP_z$, then using the fact that $P$ is a projector, which is a solution of (\ref{oureqn}),
 it is clear that
 $F$ satisfies the following properties
\bea
F_\zb=0\label{p1}\\
PF=F\label{p2}\\
FP=0.\label{p3}
\eea
Taking (\ref{p2}) with (\ref{p3}) gives that $F^2=0$, so that it has at most rank one
and can be written as 
\be 
F=uw^\dagger
\label{F}
\ee where $u$ and $w$ are two orthogonal column vectors, that
is $w^\dagger u=0.$ Substituting this expression for $F$ into (\ref{p1}) and multiplying
both sides by $w$ leads to the result that $w^\dagger u_\zb=0$, that is, $u_\zb$ is orthogonal
to $w.$ But since we already know that $u$ is orthogonal to $w$ and they are elements of
a two-dimensional vector space then this implies that $u$ and $u_\zb$ are parallel. Thus
$u$ must have the form $u(z,\zb)=g(z,\zb)\widetilde u(z)$, where $\widetilde u$ is a holomorphic
 vector and $g$ is some function.

$P$ is a Hermitian projector so it may be written as 
\be
P=\frac{vv^\dagger}{v^\dagger v}
\label{P}
\ee
for some 2-component column vector $v.$  Substituting the expressions
(\ref{F}) and (\ref{P}) into property (\ref{p2}) shows that $u$ and $v$ are
parallel, that is, $v=\lambda u$ for some function $\lambda$. Now using the
earlier factorization of $u$ we obtain $v=\lambda g \widetilde u$, so that finally
we arrive at the result that
\be
P=\frac{\widetilde u\widetilde u ^\dagger}{\widetilde u^\dagger \widetilde u}.
\ee
As we have already shown that $\widetilde u$ is a holomorphic vector then
this is an instanton solution of equation (\ref{sigma}) (see for example ref. \cite{Za})
and the required result is proved.

\end{document}